\documentclass[12pt]{article}

\makeatletter
\usepackage{epsfig}
\usepackage{amsmath}
\usepackage{setspace}
\usepackage{graphicx}

\usepackage{subfig}
\usepackage[numbers, sort&compress]{natbib}

\usepackage[utf8]{inputenc}
\usepackage[english]{babel}
\usepackage{amsmath,amssymb,amsfonts}
\usepackage{bm}
\usepackage{textcomp}
\usepackage[hmargin=2.5cm,vmargin=2.5cm]{geometry} 
\usepackage[T1]{fontenc}
\usepackage{graphicx}
\usepackage{float}
\usepackage{geometry}
\usepackage[Conny]{fncychap}
\usepackage{graphics}
\usepackage{epsfig}
\usepackage{amsmath}
\usepackage{color}
\usepackage{setspace}
\usepackage{graphicx}
\usepackage{subfig}
\usepackage[T1]{fontenc}

\usepackage{authblk}

\title{\bf Quantum Correlations Dynamics In Two Coupled Semiconductor InAs Quantum Dots}
\date{}
\author[,1,2]{\normalsize H. AIT MANSOUR\thanks{Corresponding author: \texttt{hicham.aitmansour2@gmail.com }}}
\author[1]{F-Z. SIYOURI}
\author[2]{M. FAQIR}
\author[1]{M. EL BAZ}

\affil[1]{Équipe de science de la Matière et du Rayonnement (ESMaR), 
	Faculty of Sciences, Mohammed V University, Av. Ibn Battouta, B.P. 1014, Agdal, Rabat, Morocco }
\affil[2]{International University of Rabat}
\begin{document}
	\maketitle 
	\begin{abstract}

    We investigate the dynamics of quantum discord and concurrence between two excitonic qubits placed inside two coupled semiconductor quantum dots independently interacting with dephasing reservoirs. We explore their behavior against the dimensionless time and the temperature in both Markovian and non-Markovian environments. Moreover, we analyze the external electric field effects and the Förster interaction effects on these correlations. We show that, although the quantum correlations amount is strongly influenced by the variation of the electric field and the Förster interaction, their non-Markovian behavior is still preserved under the variation of these two parameters. Furthermore, we show that for large values of temperature and dimensionless time, unlike concurrence which vanishes, nonzero discord can still be observed.

\bigskip

\smallskip

\end{abstract}

\section{INTRODUCTION}

\vspace{0.5cm}

Over the past decade, an increasing number of studies focused on investigating coupled semiconductor quantum dots\cite{inas1,inas2,inas3}, mainly because of the relevance of their electro-optical properties \cite{Nazir,Gou}.  Many different parameters are candidates for the study of these properties, among them, the exciton-exciton interaction is one of the most interesting parameters. This type of interaction in first neighbors dots will enable to carry out a scheme for quantum information processing on semiconductor quantum dots \cite{Chen}.  So, much effort has been taken to address the quantum correlation dynamics of the coupled semiconductor quantum dots. Along this direction, Fanchini et al. investigated the quantum correlation dynamics of two coupled double semiconductor quantum dots with the interaction between the qubits of the thermal bath \cite{Fanchini}. Moreover, Shojaei et al investigated the effect of external electric bias on the quantum correlations in the array of optically excited three coupled semiconductor quantum dots \cite{Shojaei}.

When it comes to quantum correlations, quantum entanglement \cite{Eins,Shro} is widely known as the key resource of quantum communication and quantum computation \cite{Niel}.  It plays an important role in the quantification of the quantum correlations present in several kinds of quantum systems \cite{Horo,Amico}. However, according to diverse studies \cite{Hend,Zurek,Zure}, entanglement cannot capture and quantify all quantum correlations. In view of this, quantum discord was introduced as a measure of all quantum correlations in \cite{Zurek,Olli}. This kind of measure is more general than quantum entanglement \cite{Olli,Mohamed} as it can exist even for some separable mixed states. In addition to that, it could be considered as a precious resource for quantum computation. For instance, studies have shown that quantum discord can speed up some quantum information tasks \cite{Datta2,Lanyon,Pira} and it was shown also that it has an immediate practical application, such as the certification of entangling gates \cite{deAl}. Furthermore, quantum discord can be very useful and can be applied in quantum metrology \cite{Girol,Adesso}.

On the other hand, in real quantum systems which are easily affected by their environments \cite{Breuer}, decoherence destroys the quantumness of the system and decreases the useful quantum correlations between the different components of the system. This decoherence is the main obstacle for the implementation of quantum information processing. It is worth noting that many studies have dealt with the dynamical behavior of quantum correlations under the effect of decoherence. Particularly the study of the quantum correlations under the influence of both Markovian and non-Markovian environments has attracted a lot of attention \cite{an,Pinto}.

 In This paper, we investigate quantum correlations between two excitonic qubits inside two coupled semiconductor quantum dots that independently interact with dephasing reservoirs. The excitonic qubits are modelled by dipoles in each quantum dots. The variation of the quantum correlations is processed as a function of temperature and the dimensionless time in both Markovian and non-Markovian environments. Our main purpose is to analyze the external electric field and the Förster interaction effect on the amount of entanglement and quantum discord measured in our system as well as on their dynamical behavior. Indeed, we show that, although the quantum correlations amount is influenced by the variation of the electric field and the Förster interaction, their non-Markovian behavior is still preserved under the effects of these two parameters. Moreover, we show that, for large values of temperature and time quantum discord still survives while entanglement vanishes.

The paper is structured as follows. In section 2, we present the Hamiltonian and the model that we use in our system. In section 3, we describe the basic concepts of concurrence and quantum discord as measures of entanglement and total quantum correlations, respectively. The results and relevant discussions are given in section 4. Finally, a brief summary is presented in section 5.

\vspace{0.5cm}
\section{Time and temperature dependent density matrix}

\vspace{0.5cm}

In order to address the quantum correlations properties of optically driven semiconductor quantum dots, we use a model sample in which we consider a series of InAs coupled semiconductor quantum dots with small equal spacing between them along the axis \cite{Expe}. In this model, to explain the energy transfer between semiconductor quantum dots through dipolar interaction between the excitons we rely on the Förster mechanism \cite{forster}. In this case, the qubits are the excitonic electric dipole moments located in each quantum dot which can only orient along ($|0\rangle $) or against ($|1\rangle$) the external electric fields.\newline
The Hamiltonian of the system in the presence of an external electric field ($\vec{E}$) is given by,
\begin{equation}
\label{eq1}
H = \hbar \sum_{i=1}^{n} \omega_i [S_z^i +\dfrac{1}{2}] + \hbar \sum_{i=1}^{n} \Omega_i S_z^i + \hbar \sum_{i,j=1}^{n} J_z [S_+^i][S_-^j]+\dfrac{1}{2} \lambda \sum_{i,j=1}^{n} [S_+^i S_-^j + S_-^j S_+^i],
\end{equation}
with $ S_+^i=\begin{pmatrix}
0 & 0 \\
1&0\\
\end{pmatrix} 
$,  $ S_-^i=\begin{pmatrix}
0 & 1 \\
0&0\\
\end{pmatrix} 
$, and  $ S_z^i=\frac{1}{2}\begin{pmatrix}
1 & 0 \\
0&-1\\
\end{pmatrix}.
$

In equation (\ref{eq1}), $\omega_i$ denotes the frequency of the excitons in the semiconductor quantum dots and $\Omega_i$ the frequency related to the excitonic dipole moment that is a function of the dipole moment and the external electric field ($\vec{E}$) at the quantum dot number $i$,
\begin{equation}
\label{eq2}
 \hbar \Omega_i= |\vec{d} .\vec{E} |,
\end{equation}
with $\vec{d} $ being the electric dipole moment associated to the exciton; it is supposed to be the same for each quantum dot. $\lambda$ is the Förster interaction which transfers an exciton from one quantum dot to another, and $J_z$ is the exciton-exciton dipolar interaction energy. For two dipoles $i$ and $j$,  along the $z$-axis and separated by a distance $r_{ij}$, it is given by
\begin{equation}
\label{eq2pr}
 \hbar J_z= \frac{\vec{d}^{2}(1-3 \cos\theta)}{\vec{r}^{3}_{ij}}.
\end{equation}

 The state of the system, when under canonical thermal equilibrium, at temperature $T$ is described by the density matrix
\begin{equation}
\label{eq3}
\rho= \frac{1}{Z}exp[- \beta H],
\end{equation}
where $\beta = ( k_BT)^{-1}$, $k_B$ being the Boltzmann constant and $T$ the temperature. The partition function  $Z$ is given by
\begin{eqnarray}
\label{eq4}
Z(T)&=&Tr( exp[- \beta H])\\
\label{eq5}
&=&\sum_{i=1}^{N} g_i e^{- \beta E_i}
\end{eqnarray}
$E_i$ being the eigenvalues of the Hamiltonian, and $g_i$ its the degeneracy.

%Therefore, the temperature-dependent density matrix of our considered system can be written as,
%\begin{equation}
%\label{eq6}
%\rho(T)=\frac{1}{Z}\sum_{i=1}^{N} e^{-\beta \lambda_i}  |\phi_i\rangle\langle\phi_i|
%\end{equation}
%here $|\phi_i\rangle$ is the eigenfunction of the Hamiltonian.\newline

 In order to evaluate the time and temperature dependent quantum correlation, we define the non-Markovian dephasing model,  introduced by Daffer et al \cite{daffer}. This model has been widely used recently for the study of quantum correlations dynamics \cite{fanchini,ali,siyouri}. In this work, we use it to address quantum correlations dynamics in double semiconductor quantum dots. We consider a colored noise dephasing model with dynamics described by the following master equation \cite{daffer,fanchini}
 \begin{equation}
 \label{eq7}
 {\dot\rho}(T) = K \mathcal{L}  \rho(T),
 \end{equation}
where the dot denotes the time derivative and $K$ is an integral operator that depends on time. It has the following form $K \phi = \int_0^t k(t- \acute{t})\phi(\acute{t})d\acute{t}$ with $k(t-\acute{t})$ being a kernel function that determines the kind of memory existing in the environment and $\mathcal{L}$ is the Lindblad super-operator describing the open system dynamics. One generally gets the master equation with Markovian approximation even in the absence of the integral operator $K$ in Eq. \eqref{eq7}. If we consider a master equation as a two-level system interacting with an environment having the properties of random telegraph signal noise, this type of master equation may arise. To analyze it, we can start with a time-dependent Hamiltonian \cite{daffer,fanchini},
 \begin{equation}
\label{eq8}
H(t) =\hbar \sum_{k=1}^3 \Gamma_k (t) \sigma_k
\end{equation}
where $\sigma_k$ are the Pauli matrices and $\Gamma_k$ (t) are independent random variables which obey the statistics of a random telegraph signal. In particular, the random variables can be given as $\Gamma_k (t)  = a_k n_k (t)$, where $n_k (t)$ has a Poisson distribution with a mean equal to $\frac{t}{2 \tau_{k}} $ and $a_k$ is an independent random variable taking values $\pm a_k$.

The equation of motion for the density matrix is giving by the von Neumann equation $ {\dot\rho}(T) = -(i/\hbar)[H, \rho(T)]= -i \sum_k \Gamma_k (t) [\sigma_k, \rho(T)]$, which admits the solution
 \begin{equation}
\label{eq9}
\rho(t,T) = \rho(0,T)- i \int_0^t \sum_k \Gamma_k (s) [\sigma_k, \rho(s,T)]ds.
\end{equation}
We substitute this equation back into the von Neumann equation and we perform a stochastic average, we obtain then,
\begin{equation}
\label{eq10}
\dot{\rho}(t,T)= -\int_0^t \sum_k \exp(-\left|t-\acute{t}\right|/\tau_{k}) a_k^2 [\sigma_k,[\sigma_k,\rho(\acute{t},T)]] d\acute{t}
\end{equation}
where $< \Gamma_j (t)\Gamma_k (\acute{t})> =  a_{k}^{2} \exp(-\left|t-\acute{t}\right|/\tau_{k}) \delta_{jk}
$ is the memory kernel obtained from the correlation functions of random telegraph signal.\newline
The conditions in which the dynamical evolution created by Eq. (9) is wholly non-negative has been analysed by Daffer et al. \cite{daffer} and they found that complete positivity is guaranteed when two of the $a_k$'s are zero. This is consistent with a physical situation where the noise only acts in one direction. Particularly, provided that the condition $a_1=a_2=0$, and $a_3=a$ holds, the dynamics experienced by the system is that of a dephasing channel with colored noise. Hence, the Kraus operators that describe the dynamics of the two-level system are given by \cite{daffer,fanchini}, 
 \begin{equation}
 \label{eq11}
 M_1 = \sqrt{\frac{1 + \Lambda(\nu)}{2}} I_2 ,
 \end{equation}
 
 \begin{equation}
 \label{eq12}
 M_2 = \sqrt{\frac{1 - \Lambda(\nu)}{2}} \sigma_3
 \end{equation}
where $I_2$ is the $ (2\times 2)$ identity matrix and the Kraus operators $ M_i (t)$ satisfy $\sum_i M_i^{\dagger} (t) M_i (t) = I$. Here,  $\Lambda(\nu)=e^{- \nu}[\cos(\mu \nu) +\sin(\mu \nu)/\mu]$, with $\mu= \sqrt{(4 a \tau)^2 -1}$ and $\nu = \frac{t}{2\tau}$ is the dimensionless time.

Since we intend to study the dynamics of a two, two-level quantum system (two quantum dots), the time evolution of an initial density operator, $\rho_{AB}(0,T)$, can be expressed as 
\begin{equation}
\label{eq13}
\rho_{AB}(t,T)=\sum_{i,j}(M_i^A\otimes M_j^B)\rho_{AB}(0,T)(M_i^A\otimes M_j^B)^{\dagger}
\end{equation}
where the Kraus operators $M_i^A$ and  $M_i^B$ act, respectively, on the first and the second qubit.

The time and temperature dependent density matrix is then expressed in the following form,
\begin{equation}
\label{eq14}
\rho_{AB}(t,T) =\nonumber \begin{pmatrix} 
\rho_{11 }& 0  & 0 &0\\

0 & \rho_{22} & \rho_{23} & 0 \\

0 &\rho_{32 }& \rho_{33 }& 0 \\

0 & 0 & 0  & \rho_{44 } 

\end{pmatrix} 
\end{equation}
\begin{scriptsize}
with $$\rho_{11 }= 
\dfrac{e^{- \beta \hbar(\omega+\Omega)}}{2(Cosh(\beta \sqrt{\hbar J_z (\lambda+\hbar J_z)})+Cosh(\beta \hbar (\omega+\Omega)))},$$

$$\rho_{22 }= \dfrac{\hbar J_z}{(\lambda + \hbar J_z)(1+Cosh(\beta \hbar(\omega+ \Omega)) Sech(\beta \sqrt{\hbar J_z (\lambda + \hbar J_z)}))}, $$ 
$$
 \rho_{32 }= - \dfrac{e^{-2\nu}\hbar J_z (8 a^2 \tau^2 +(-1 + 8a^2 \tau^2)Cos(2\nu \sqrt{-1+16 a^2 \tau^2})+\sqrt{-1 + 16 a^2 \tau^2} Sin(2\nu\sqrt{-1+16 a^2 \tau^2 }))Sinh(\beta \sqrt{\hbar J_z(\lambda + \hbar J_z)})}{(-1+16 a^2 \tau^2 )\sqrt{\hbar J_z (\lambda+\hbar J_z)}(Cosh(\beta \sqrt{\hbar J_z(\lambda +\hbar J_z)})+Cosh(\beta \hbar(\omega+\Omega)))},$$
$$
\rho_{23 }=- \dfrac{e^{-2\nu}\hbar J_z (8 a^2 \tau^2 +(-1+8a^2\tau^2)Cos(2\nu \sqrt{-1+16a^2\tau^2})+\sqrt{-1+16a^2 \tau^2}Sin(2\nu \sqrt{-1+16 a^2 \tau^2}))Sinh(\beta \sqrt{\hbar J_z(\lambda+\hbar J_z)})}{(-1+16 a^2 \tau^2)\sqrt{\hbar J_z (\lambda + \hbar J_z)}(Cosh(\beta \sqrt{\hbar J_z(\lambda+\hbar J_z)}))+Cosh(\beta \hbar(\omega+\Omega))},
 $$
$$
\rho_{33 }= \dfrac{1}{1+Cosh(\beta \hbar (\omega+\Omega))Sech(\beta \sqrt{\hbar J_z (\lambda+\hbar J_z)})},
$$

and $$\rho_{44 }= \dfrac{e^{\beta \hbar(\omega +\Omega)}}{2(Cosh(\beta \sqrt{\hbar J_z(\lambda +\hbar J_z)})+Cosh(\beta \hbar(\omega+\Omega)))}.$$

\end{scriptsize}
\vspace{0.5cm}

\section{Quantum correlations}
\vspace{0.5cm}
\subsection{Concurrence}

The concurrence \cite{wootters} is one of the most frequently used measures of entanglement of a general two-qubit system. The concurrence of a bipartite system $AB$ composed of the subsystems $A$ and $B$ and the state of which is described by $\rho_{AB}$, is defined by,
\begin{equation}
\label{eq15}
C(\rho_{AB})=Max[0,(\lambda_1 -\lambda_2 -\lambda_3 -\lambda_4  )]
\end{equation}
where the $\lambda_i$'s are the square roots of the positive eigenvalues of the matrix $ \rho_{AB}.\tilde{\rho}_{AB}$ arranged in decreasing order. The spin flipped density matrix $\tilde{\rho}_{AB}(t,T)$ is defined by,
\begin{equation}
\label{eq16}
\tilde{\rho}_{AB}=(\sigma_y \otimes \sigma_y)\rho_{AB}^{\ast}  (\sigma_y \otimes \sigma_y)\,
\end{equation}
with $\sigma_y $ being the Pauli matrix and $\rho_{AB}^{\ast}$, the complex conjugate of $\rho_{AB}$.

\subsection{Quantum Discord}

The quantum discord \cite{discord} is a measure of the non-classical correlations between two subsystems of a quantum system. For a bipartite system $AB$, it is defined as the difference between two alternative quantum versions of two classically equivalent expressions of the mutual information:
\begin{equation}
\label{eq17}
D(\rho_{AB})=I(\rho_{AB})-C(\rho_{AB}),
\end{equation} 
where $I(\rho_{AB})$ and $C(\rho_{AB})$  are, respectively, the mutual information and the classical correlations of the composite system $\rho_{AB}$. They are defined, respectively by 
\begin{equation}
\label{eq18}
I(\rho_{AB})=S(\rho_A)+S(\rho_B)-S(\rho_{AB})
\end{equation} 
and 
\begin{equation}
\label{eq19}
C(\rho_{AB})=Max[S(\rho_A)-S(\rho_{AB}/{\left\{ \Pi_{B}^{j}\right\}})].
\end{equation}
The von Neumann entropy, $S(\rho) $, \cite{nielsen} is defined by $ S(\rho)=-Tr(\rho \log_2 \rho) $ and   $\rho_A , \rho_B $ are the reduced density matrices of the composite system $\rho_{AB}$. In (\ref{eq19}) the conditional entropy is given by $S(\rho_{AB}/{\left\{ \Pi_{B}^{j}\right\}})=\sum_j P_j S(\rho_{A/j})$ with $\/{\left\{ \Pi_{B}^{j}\right\}}  $ being the complete set of orthonormal projection operators that act only on the second subsystem $B$, $\rho_{A/j}=Tr( \dfrac{\Pi_B^j \rho_{AB}  \Pi_B^j}{P_j} )    )$ is the  resulting state of the first subsystem $A$, and $P_j = Tr_{AB} ( \Pi_B^j \rho_{AB} \Pi_B^j)  $  is the probability to obtain the outcome $j$.

The set of local measurement operators $\left\{\prod_B^{j}\right\}\equiv\left\{{|\Pi_1 \rangle \langle \Pi_1 |, |\Pi_2 \rangle \langle \Pi_2 |}\right\}$ can be easily constructed using the following states: 
$| \Pi_1 \rangle = \cos \theta| + \rangle + e^{i \Phi}\sin \theta |- \rangle$  \quad \hbox{and} \quad  $| \Pi_2 \rangle= \sin \theta |+  \rangle- e^{i \Phi}\cos \theta |- \rangle$, with $\theta\in [0,\pi] $ and $\Phi \in [0,2\pi]$.

Therefore, the quantum discord capturing all quantum correlation can be rewritten as \cite{henderson}
\begin{equation}
 \label{eq2021}
D(\rho_{AB}) =S(\rho_B)-S(\rho_{AB})  +  Min_{               \left\lbrace \Pi_B^j      \right\rbrace        }[S(\rho_{AB}/{\left\{ \Pi_{B}^{j}\right\}} ].
 \end{equation}

It turns out that the conditional entropy minimization is the main difficulty in finding an analytic expressions for quantum discord present in arbitrary states. Indeed, the procedure of calculation of this later is not easy for all states and the exact analytical expressions are found only in a limited number of cases and the most general approach until now was obtained for the so-called X-states.  Here, the density matrix of our system is written in the form of these states. In fact, the method given by C.Z Wang et al \cite{discx} can be used to calculate the quantum discord. As a matter of fact, this later can be redefined by the following expression, for $X$-states \cite{discx,discxa}
\begin{equation}
\label{eq22}
 D(\rho_{AB})=Min[QD_1,QD_2],
\end{equation} 
where  $$ QD_j=H(\rho_{11 }+\rho_{33 }) +\sum_{k=1}^{4} \lambda_k Log_2(\lambda_k)+Dj.$$

Here $ D_1=H(\eta)$ and $D_2=-\sum_{j=1}^{4}\rho_{jj} Log_2(\rho_{jj}) -H(\rho_{11 }+\rho{33})$, with $ \eta=\frac{1+\sqrt{[1-2(\rho_{33 }+\rho_{44 })]^2+4 (|\rho_{14}|+|\rho_{23}|)^2}}{2}$, $ H(x)=-x Log_2(x)-(1-x)Log_2(1-x)$ is the binary Shannon entropy and the $\lambda_k$'s are the eigenvalues of the matrix $\rho_{AB}$.

\vspace{0.5cm}
\section{ Results and discussions}

We investigate the behavior of the quantum correlations present in two coupled semiconductor quantum dots independently interacting with dephasing reservoirs. In figure $\ref{fig1}$ we plotted the evolution of both concurrence and discord for different values of $\tau$ against the parameter $\frac{t}{2 \tau}$ when the electric field is zero ($\ref{fig1}$ (a, b,d,e)) and also when it is equal to $ 25\times 10^6 $V/m ($\ref{fig1}$ (c,f)). Moreover, in order to process the effects of Förster interaction on the correlations present in our system we plotted in figure \ref{fig2} their dynamical behavior for different values of this parameter in non-Markovian regime. Note that in all cases we fix $a=0.9s$ and $\lambda = $. \newline  

 \begin{figure}[H]
 \centering
 	\subfloat[][E: $ \hbar \Omega = 0 V m^{-1}$ and $ T=14 K$]{\includegraphics[width=0.33\textwidth]{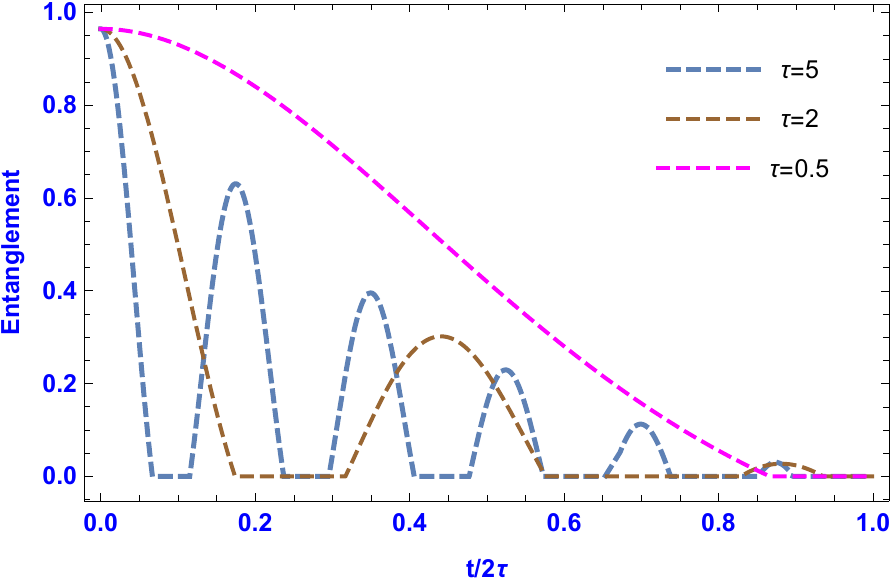}\label{fig1a}}
	\subfloat[][E: $ \hbar \Omega = 0 V m^{-1}$ and $ T=25 K$]{\includegraphics[width=0.33\textwidth]{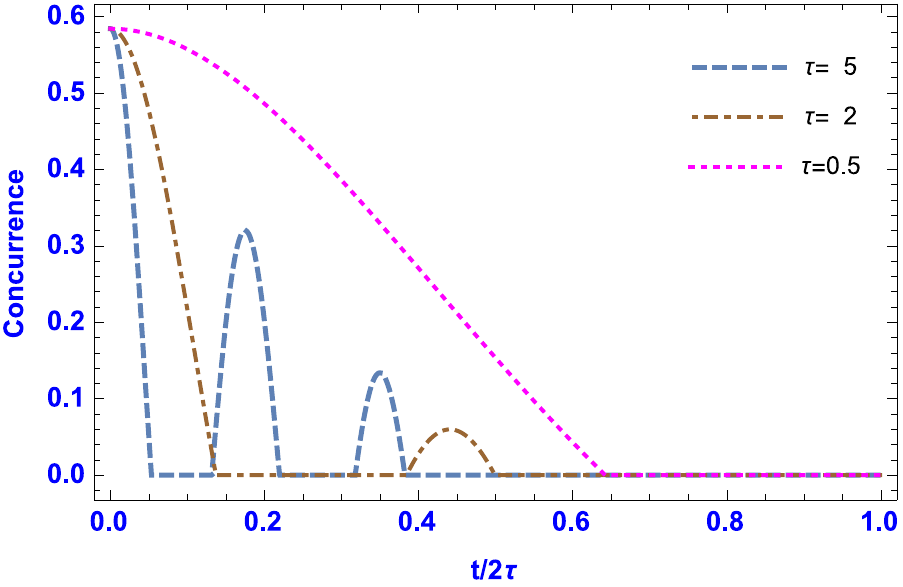}\label{fig1b}}
	\subfloat[][E: $\hbar \Omega =25\times 10^6 V m^{-1}$ and $ T=25 K$]{\includegraphics[width=0.33\textwidth]{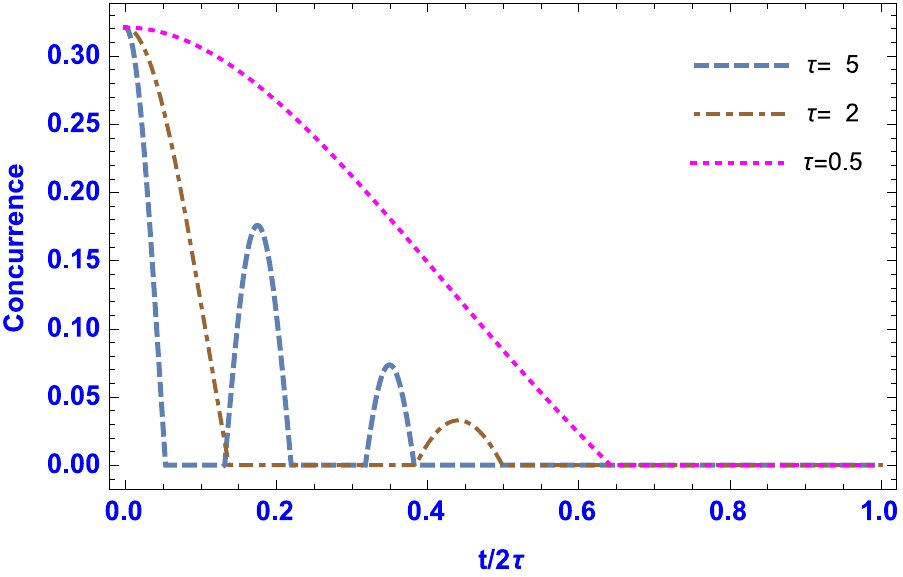}\label{fig1c}}\\
	\subfloat[][D: $ \hbar \Omega = 0 V m^{-1}$ and $ T=14 K$]{\includegraphics[width=0.33\textwidth]{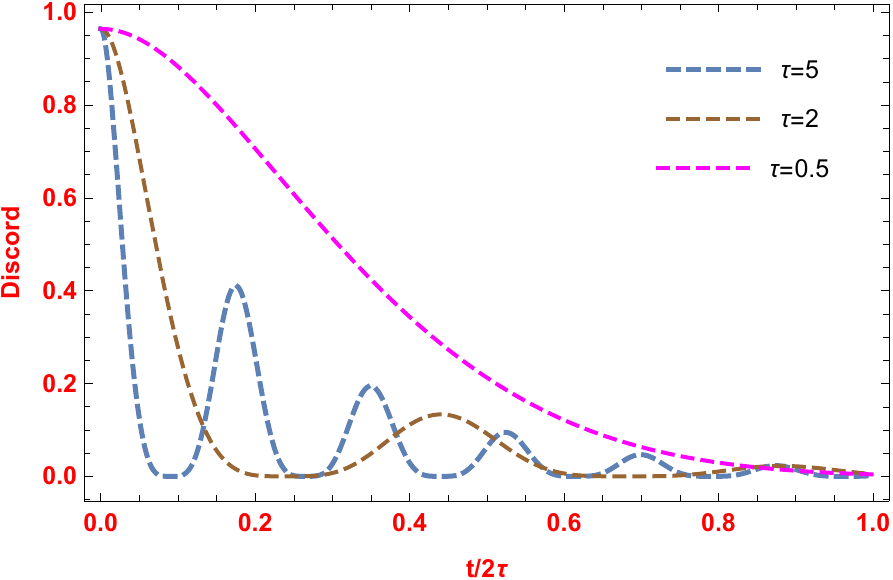}\label{fig1d}}
	\subfloat[][D: $ \hbar \Omega = 0 V m^{-1}$ and $ T=25 K$]{\includegraphics[width=0.33\textwidth]{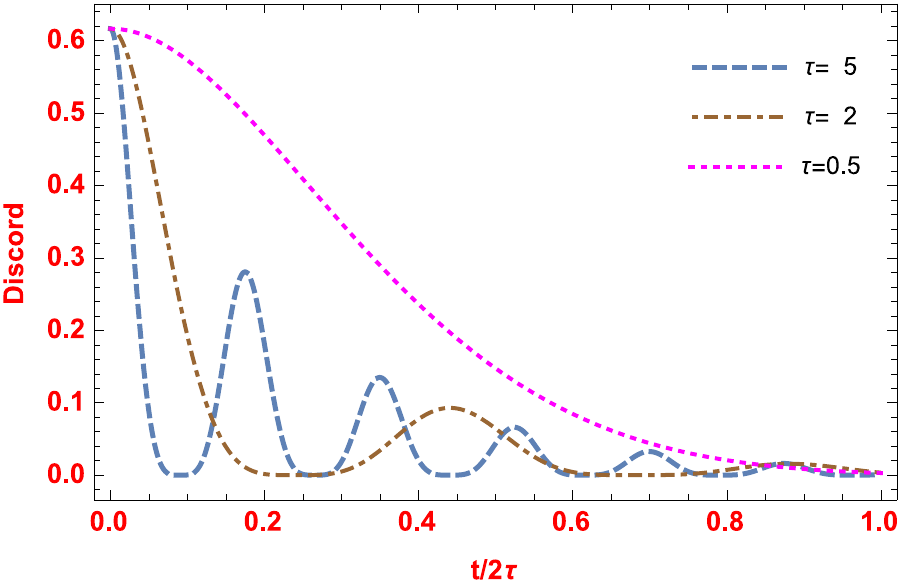}\label{fig1e}}
	\subfloat[][D: $\hbar \Omega =25\times 10^6 V m^{-1}$ and $ T=25 K$]{\includegraphics[width=0.33\textwidth]{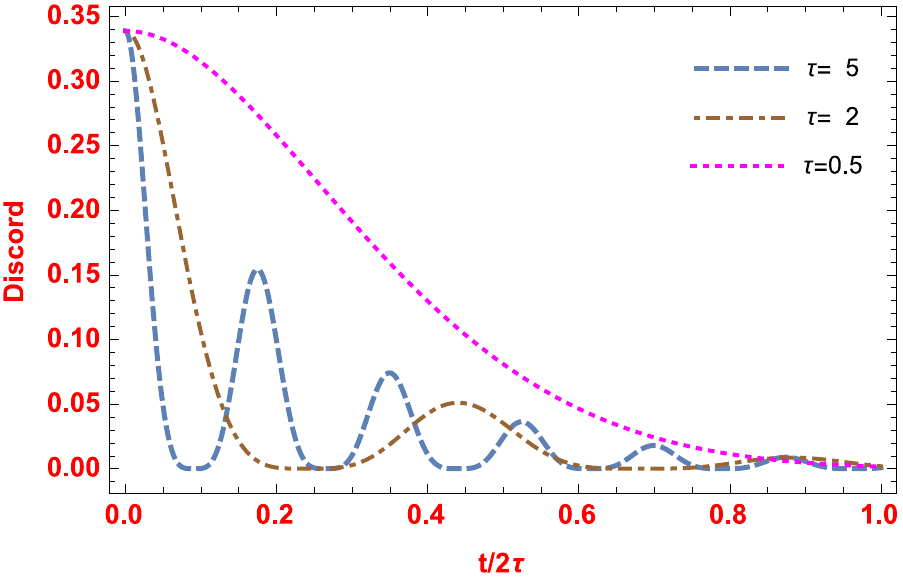}\label{fig1f}}
	\caption{\label{fig1}Entanglement ($E$) and Discord ($D$)  as a function of $\frac{t}{2 \tau}$ for different values of $\tau$ with  $\lambda=0$ }.
	\label{fig1}
\end{figure}

Figures \ref{fig1a}, \ref{fig1b}, \ref{fig1d}, and \ref{fig1e} show how quantum correlations behave against the dimensionless time for two different values of temperatures when the electric field is absent. For $ \tau \geq 2$ (i.e. the non-Markovian regime), one can observe that the quantum correlations (both entanglement and quantum discord) exhibit death and revival with a continuous damped amplitude. This is due to the environment memory effects which lead to the increase in the information back-flow. While for $\tau=0.5$ (i.e. the Markovian regime), correlations are decreasing asymptotically to zero with increasing time without any revival. This behavior is explained by the fact that the quantum information will very quickly outflow from the system to the environment because of the weak system-environment coupling and the memoryless transfer of information. Moreover, one can observe that whenever the degree of non-Markovianity $ \tau $ decreases, the frequency of the oscillations in the correlation function is also decreased and delayed, as well as the amplitude of this correlation function. Furthermore, it is seen that, the amount of quantum correlations decreases once we increase the temperature (Figures \ref{fig1b} and \ref{fig1e}).  Additionally, one can remark that for large values of dimensionless time and at higher temperature $(T=25 K)$ nonzero discord can still be observed, unlike concurrence which vanishes. This shows the relevance of quantum discord in measuring the quantum correlations. On the other hand,  Figures \ref{fig1c} and \ref{fig1f} show that when the applied electric field is on, quantum correlations behave in a similar manner as the plots of Figures \ref{fig1b} and \ref{fig1e}, however with smaller amplitudes than those obtained when the electric field is off. This is due to tendency of all dipoles to align in one direction when the external electric field is applied. this results in an increasing dipole-dipole repulsive interaction that forthwith causes the reduction of Coulomb induced correlations. 

\begin{figure}[H]
	\centering
	\includegraphics[width=0.48\linewidth]{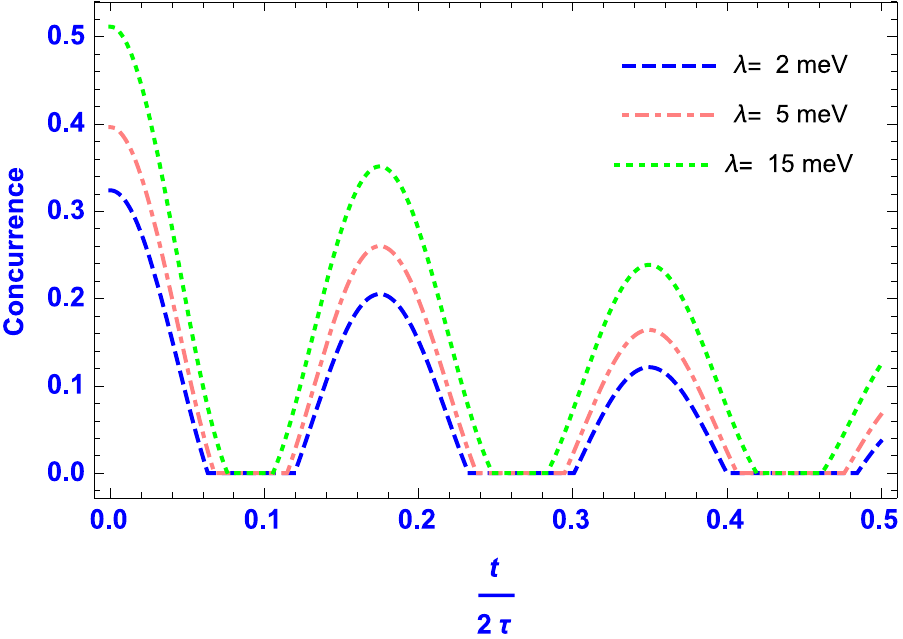}\label{fig2a}
	\includegraphics[width=0.48\linewidth]{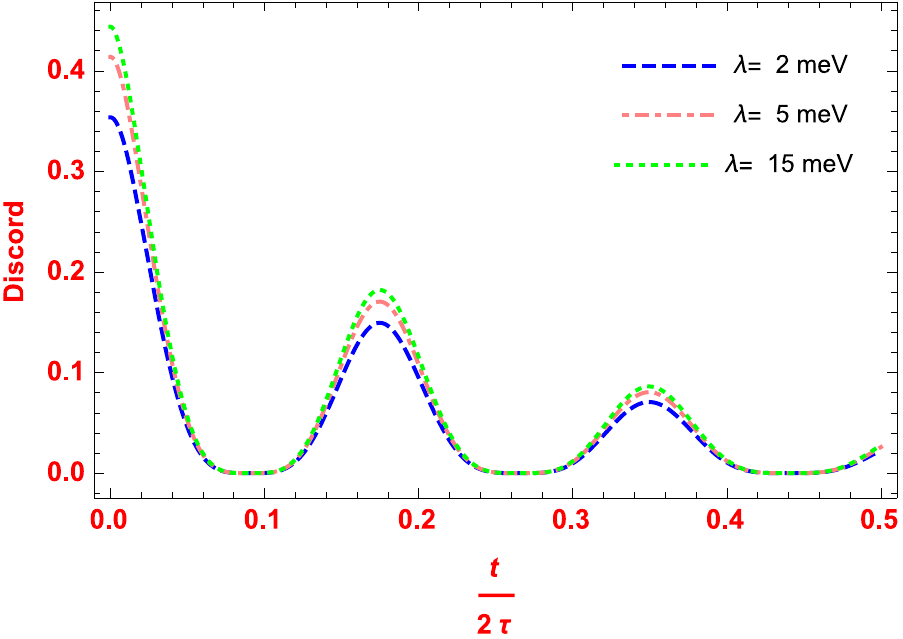}\label{fig2b}
	\caption{\label{fig2}Entanglement and Discord  as a function of $\frac{t}{2 \tau}$ for different values of $\lambda$, described by the parameters $\tau=5$, $ T=20 K$ and $ \hbar \Omega = 30\times 10^6 V m^{-1}$}
\end{figure}

In Figure \ref{fig2} which displays the dynamics of quantum correlations for different values of Förster interaction $\lambda$ when the electric field is on ( $\hbar \Omega =30\times 10^6 V m^{-1}$), one can observe that the amount of quantum correlations increases with $\lambda$. This increase can be resulted from the increasing excitonic interaction. It is worthwhile noting that the quantum correlations preserve their non-Markovian behavior under the Förster interaction effects and the electric field effects as obsereved in Figure \ref{fig1}.

\begin{figure}[H]
 	\centering
 	 \includegraphics[width=0.48\linewidth]{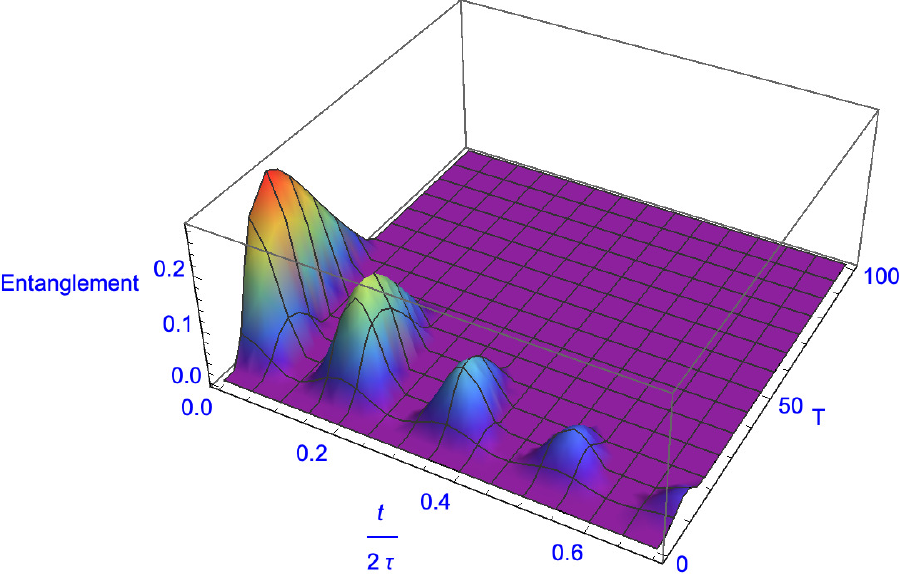}\label{fig3a}
 	\includegraphics[width=0.48\linewidth]{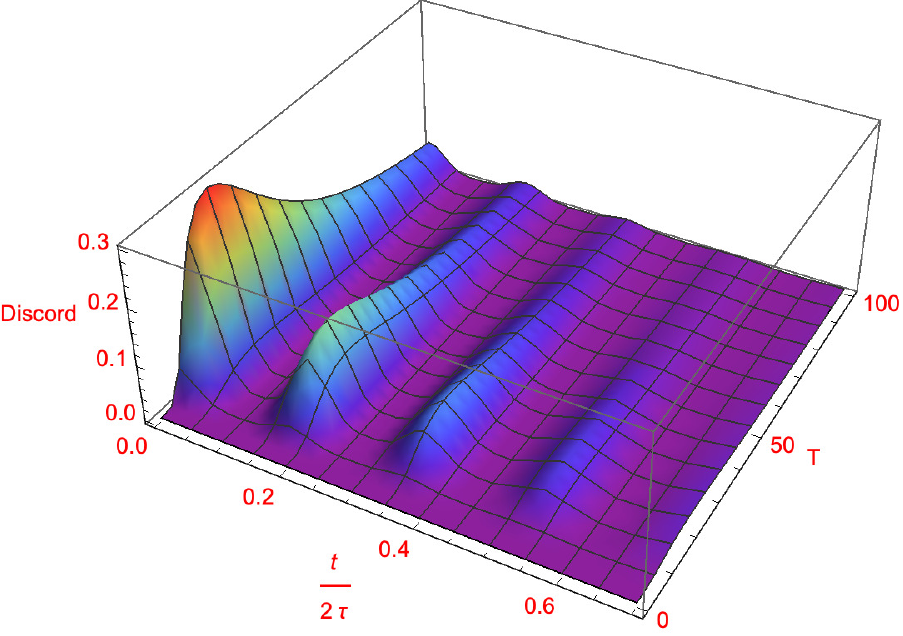}\label{fig3b}
 	\caption{\label{fig3} Entanglement and Discord with respect to $T$ and $ \dfrac{t}{2\tau}$, for $\tau=5$, $\lambda=0$ and $\hbar \Omega =25\times 10^6 V m^{-1}$}
 \end{figure}

\begin{figure}[H]
	\centering
	\includegraphics[width=0.48\linewidth]{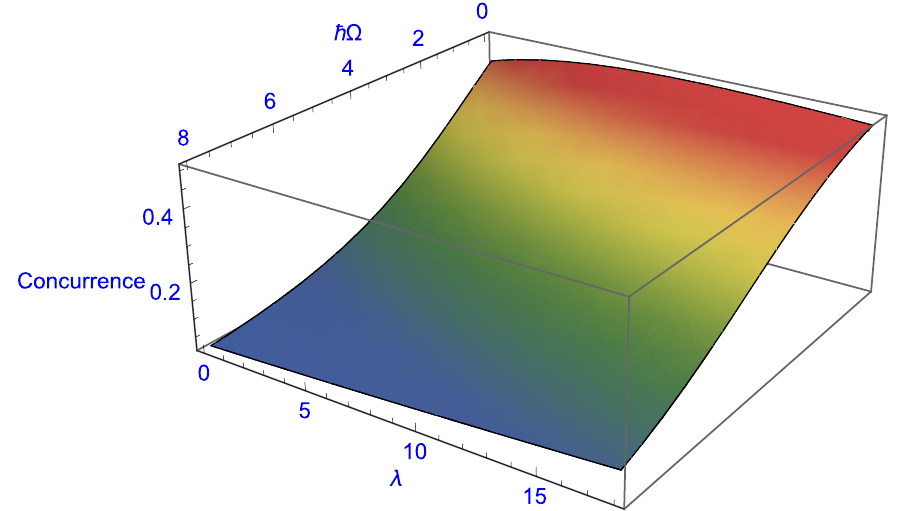}\label{fig4a}
	\includegraphics[width=0.48\linewidth]{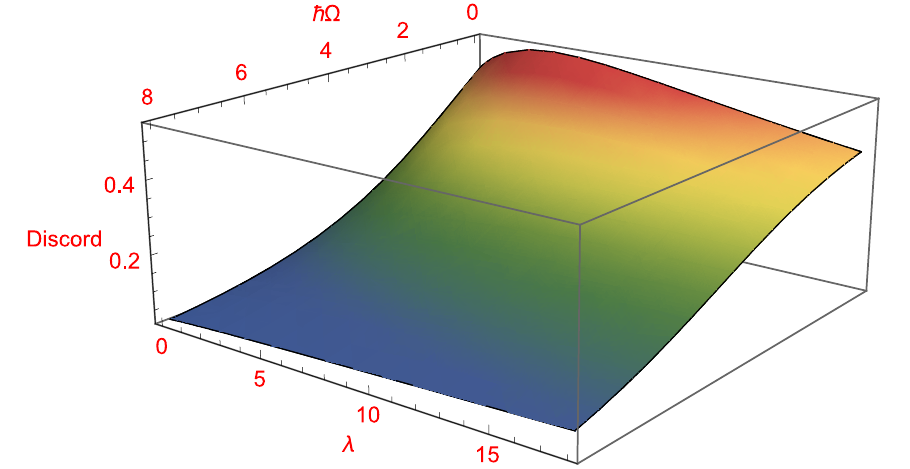}\label{fig4b}
	\caption{\label{fig4} Entanglement and Discord as a function of $\lambda$ and  $\hbar \Omega $, for $\tau=5$, $T=25K$ and $ \dfrac{t}{2\tau}=0.01$}
\end{figure}

Now, in order to study the evolution of quantum correlations more rigorously, we plotted their 3D dynamical behavior  against the dimensionless time  $\frac{t}{2 \tau}$ and the temperature $T$ in Fig. \ref{fig3} as well as against the Förster interaction $\lambda$ and the external electric field $\hbar \Omega$  in Fig. \ref{fig4} in the non-Markovian case; $\tau =5 $.

It is clearly seen that both concurrence and discord decay with temperature and dimensionless time while exhibiting death and revival. This decay is due to the thermal relaxation effects and environment effects, respectively. For smaller temperatures, one can remark that both quantum discord and concurrence increase until reaching their maximum values then decrease gradually with temperature without any revival as expected. However, one can remark that for higher values of temperature discord still survives while concurrence vanishes. As a matter of fact, one can again assert that quantum discord is more robust than concurrence against temperature. On the other hand, As it is seen in Figure \ref{fig1} one can observe that quantum correlations behavior against time presents some collapses and revivals because of the environment memory effects.

The plots in Figure \ref{fig4} show that whenever we increase the electric field the amount of both quantum discord and concurrence decreases. This decrease can be explained in terms of the increasing dipolar repulsive interaction as all the dipoles become parallel under effect of the electric field. Moreover, the figures show that for larger values of the electric field these correlations increase with the Förster interaction. Whereas, for very low values of the electric field the quantum discord diminishes slightly at higher Förster interactions in contrast to the concurrence which reaches a constant value. In fact, one can state that increasing the electric field perturbs the effect of Förster interactions on quantum correlations.

\section{CONCLUSION}

In this work, we have investigated the variation of quantum discord and entanglement in the array of two optically driven coupled semiconductor quantum dots independently interacting with dephasing reservoirs.  Each quantum dot has an exciton that can be modelled by an electric dipole.  

We have shown that, the amount of quantum correlations increases upon increasing the Förster interaction and diminishes whenever we switch the electric field on. Moreover, we have shown that, this amount increases and reaches its maximal value at very low temperatures and dimensionless time regions then decreases and decays with increasing these two parameters due to the thermal relaxation effects and environment effects. However, we have observed that despite the fact that concurrence vanishes for large values of temperature  and dimensionless time quantum discord still survives, which is consistent with the fact that discord quantifies quantum correlations beyond entanglement. In this direction, we have concluded that, although the quantum correlations amount is influenced by the electric field effects and the Förster interaction effects, their non-Markovian behavior is still preserved under these two effects.

\end{document}